\documentclass[prl, twocolumn, floatfix, superscriptaddress, showpacs, longbibliography]{revtex4-2}

\pdfoutput=1						
\usepackage[charter]{mathdesign}	
\usepackage{amsmath}				

\usepackage{mathrsfs}					

\usepackage{mathtools}
\usepackage[]{stackengine}

\newlength\correct
\usepackage[pdftex]{graphicx}
\usepackage{microtype}
\usepackage{dsfont}
\usepackage{bm}
\usepackage{gensymb}
\usepackage{mdframed}
\usepackage[caption=false]{subfig}

\usepackage[svgnames]{xcolor}
\usepackage[
	colorlinks=True,linkcolor=DarkRed,citecolor=ForestGreen,urlcolor=MediumBlue,
	pdfstartview=FitH,bookmarks=False,pdfpagemode=UseNone
]{hyperref}

\usepackage{feynmp-auto}
\usepackage{natbib}
\usepackage{hyperref}
\usepackage{float}
\usepackage{cancel}

\usepackage[T1]{fontenc}
\usepackage{inputenc}
\usepackage{tcolorbox}
\usepackage{stackengine}

\definecolor{KIT-green}{RGB}{0, 150,130}
\definecolor{KIT-blue}{RGB}{70,100,170}

\makeatletter
\let\cat@comma@active\@empty
\makeatother

\begin{document}

\title{BCS to incoherent superconductivity crossovers in the Yukawa-SYK model on a lattice}
\author{D. Valentinis}
\affiliation{Institut f\"{u}r Quantenmaterialien und Technologien, Karlsruher Institut
f\"{u}r Technologie, 76131 Karlsruhe, Germany}
\affiliation{Institut f\"{u}r Theorie der Kondensierten Materie, Karlsruher Institut
f\"{u}r Technologie, 76131 Karlsruhe, Germany}
\author{G. A. Inkof}
\affiliation{Institut f\"{u}r Theorie der Kondensierten Materie, Karlsruher Institut
f\"{u}r Technologie, 76131 Karlsruhe, Germany}
\author{J. Schmalian}
\affiliation{Institut f\"{u}r Quantenmaterialien und Technologien, Karlsruher Institut
f\"{u}r Technologie, 76131 Karlsruhe, Germany}
\affiliation{Institut f\"{u}r Theorie der Kondensierten Materie, Karlsruher Institut
f\"{u}r Technologie, 76131 Karlsruhe, Germany}

\date{\today}

\begin{abstract}
We provide a quantitative and controlled analysis of the phase diagram of the the Yukawa-SYK model on a lattice, in the normal and superconducting states.  We analyze the entire crossover from BCS/weak-coupling to Eliashberg/strong coupling superconductivity, as a function of fermion-boson interaction strength and hopping parameter. Cooper pairs of sharp Fermi-liquid quasiparticles at weak coupling evolve into pairing of fully incoherent fermions in the non-Fermi liquid regime. The crossovers leave observable traces in the critical temperature, the zero-temperature and zero-energy gap, the entropy, and the phase stiffness. 
\end{abstract}

\maketitle
For an understanding of superconductivity, a clear grasp of the corresponding normal state is essential, both near the transition temperature where it is the normal state that becomes unstable, and deep in the paired state when the energy scales of the superconducting phase are small \cite{Emery-1994, Emery-1995}. As we will see, such grasp is equally important when all scales are of comparable magnitude.
\begin{figure*}[t]
\includegraphics[width=0.9\textwidth]{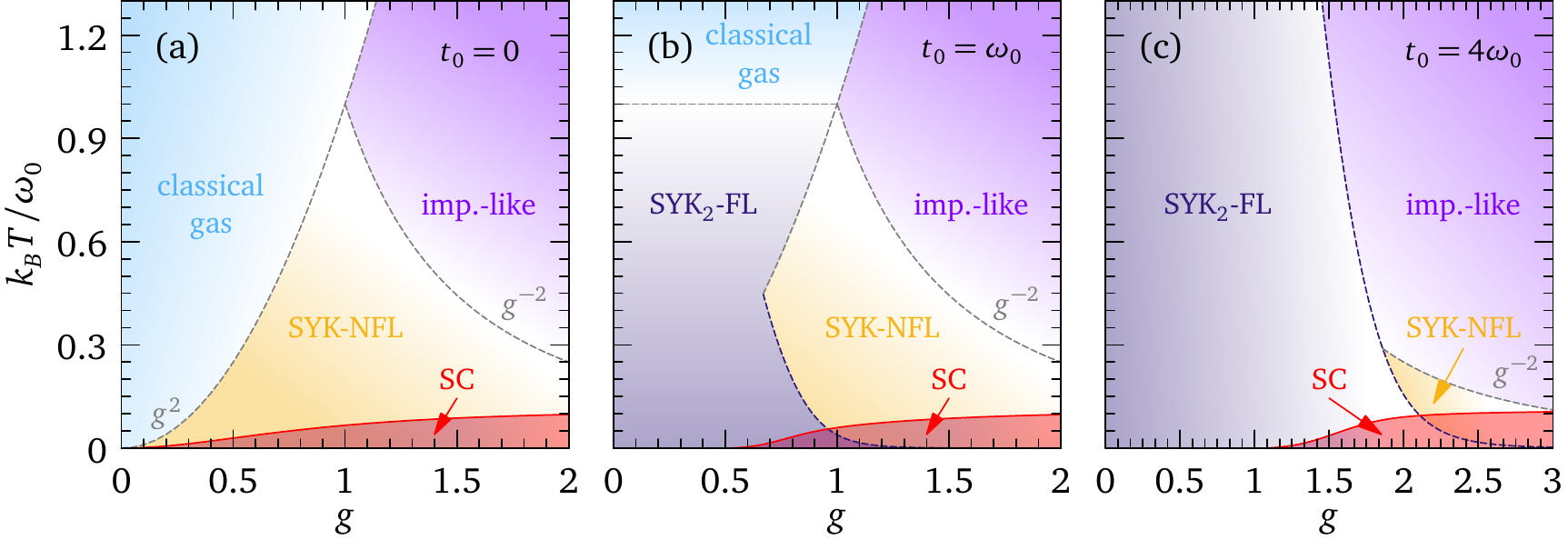}
\caption{\label{fig:crossovers} Phase diagrams of the Yukawa-SYK model on a lattice, as a function of dimensionless fermion-boson interaction $g$ and temperature $k_B T/\omega_0$, for different hopping energies: (a) $t_0=0$; (b) $t_0=\omega_0$; (c) $t_0=4 \omega_0$. }
\end{figure*}

A crucial assumption about the normal state made by the Bardeen-Cooper-Schrieffer (BCS) theory of superconductivity \cite{Bardeen-1957a,Bardeen-1957b,Schrieffer-1963th} is that it can be described as a Fermi liquid (FL) \cite{Abrikosov-1959,Abrikosov-1972FL}. For conventional superconductors this assumption is well justified. The Cooper instability, i.e.\@, the formation of a two-particle bound state at arbitrarily small attractive interaction, relies crucially on the normal state being a FL \cite{Cooper-1956}.

On the other hand, in correlated electron systems superconductivity is frequently observed when FL theory does not seem to apply \cite{Mathur-1998,Petrovic-2001,Sidorov-2002,Nakatsuji-2008,Knebel-2011,Kasahara-2010,Bohmer-2014,Shibauchi-2014,Kuo-2016, Zaanen-2004,Valla-1999,Bruin-2013,Giraldo-2018,Naqib-2019,Mandal-2019,Michon-2019,Legros-2019,Grissonnanche-2021,Keimer-2015}. This can, for example, be due to soft collective modes that behave at small but finite temperatures like almost classical degrees of freedom, possibly with additional quantum fluctuations. Then, single-particle momentum states rapidly broaden at temperatures much smaller than the bandwidth, even right at the Fermi energy and - depending on the pairing state and collective modes involved - pair-breaking phenomena weaken the Cooper instability. A related but different reason for the failure of FL-theory is the vicinity to a  quantum critical point (QCP). Here, slow long-range interactions, mediated by critical modes, give rise to a singular single-particle scattering rate, even as $T\rightarrow 0$. Then the usual Cooper instability is absent. It is rather surprising that for a number of materials the superconducting transition temperature is largest right at a QCP \cite{Mathur-1998,Petrovic-2001,Sidorov-2002,Nakatsuji-2008,Knebel-2011,Kasahara-2010,Bohmer-2014,Shibauchi-2014,Kuo-2016}. In addition, even when born out of an incoherent normal state, systems like the cuprate superconductors display coherent and dispersing Bogoliubov quasiparticles below $T_c$ \cite{Dessau-1991,Campuzano-1996,Loeser-1997,Shen-1997,Fedorov-1999,Kaminski-2016,Orenstein-2000,Hashimoto-2014}. The unusual nature of the normal state is reflected in a small weight of these coherent quasiparticles.
In Refs.\@ \onlinecite{Shen-1997,Feng-2000} a close connection between the spectral weight of Bogoliubov quasiparticles, the superfluid stiffness, and the condensation energy was reported for underdoped cuprates.
    
In this paper we present a  controlled theory that describes pairing of a system with and without quasi-particles and is able to describe the crossover between the two regimes. This enables us to establish several trends of how an unconventional normal state affects the superconducting phase, trends that might be more general than the model we use to derive them.
We use a lattice version of the Yukawa-SYK model of Ref.\@ \onlinecite{Esterlis-2019}, which considers isolated quantum dots with a large number of mutually interacting internal degrees of freedom \cite{Esterlis-2019,Inkof-2022,Hauck-2020,Wang2020b,Wang-2020a,Classen-2022,Choi-2022}. 
Coupling these dots with a single-particle hopping amplitude yields, in the normal state, a crossover from quantum critical regime at high energies to a disordered FL at low energies \cite{Song-2017,Guo-2022, Chowdhury-2022,Patel-2023_preprint}. Depending on the pairing strength, superconductivity emerges in the quantum-critical and the FL regimes. Our key findings are that $T_c$ is largest at the quantum critical point and decreases as one moves towards a FL regime. This is despite the fact that the usual Cooper instability is no longer active: it is replaced by a more singular pairing interaction at criticality that is more efficient than pairing in FLs \cite{Bonesteel-1996,Son-1999,Abanov-2001a,Abanov-2001b,Chubukov-2005,She-2009,Moon-2010,Metlitski-2015,Roussev-2001,Raghu-2015}. At the same time we observe that the stiffness is largest at the crossover from FL to non-Fermi liquid (NFL) behavior. The stiffness in the disordered FL is small because of the comparative weakness of Cooper pairing, while it is small deep in the incoherent regime because of the small optical weight of the normal state. In between these regimes the stiffness is largest.
Deep in the NFL regime we further observe a correlation between the Bogoliubov quasiparticle weight, the condensation energy and the stiffness, in qualitative agreement with observations \cite{Shen-1997,Feng-2000}. While our model is clearly motivated by these classic observations, we do not pretend that it can be directly applied to systems as complex and diverse as the cuprate, organic, heavy-fermion, or iron-based superconductors. Nevertheless, we suspect that some of the findings of our analysis are of more general in character and help develop a broader phenomenology of pairing in systems without quasiparticles.

We consider a model where Yukawa-SYK dots are coupled by a random single particle hopping. This is analogous to coupled purely fermionic SYK models \cite{Song-2017,Can-2019, Salvati-2021,Chowdhury-2022}, but it also allows for superconducting solutions due to coupling between fermions and a scalar bosonic field. The model is given as
 The model is given as
 \begin{equation}
    \hat{H}=\frac{1}{\mathscr{N}}\sum_{ij}\sum_{\left\langle \boldsymbol{x},\boldsymbol{x}'\right\rangle }t_{ij,\boldsymbol{x} \boldsymbol{x}'}\hat{c}_{i,\boldsymbol{x},\sigma}^{\dagger}\hat{c}_{j,\boldsymbol{x}',\sigma}+\sum_{\boldsymbol{x}}\hat{H}_{\boldsymbol{x}}.
\end{equation} 
$\hat{c}_{i,\boldsymbol{x},\sigma}$ and $\hat{c}_{i,\boldsymbol{x},\sigma}^{\dagger}$ are fermionic annihilation operators at site $\boldsymbol{x}$, flavor index $i$ and with spin $\sigma$. $t_{ij,\boldsymbol{x}\boldsymbol{x}'}$ is a random single-particle hopping with mean square value $t_0/\sqrt{z}$ ($z$ is the coordination number), acting between neighboring sites $\boldsymbol{x}$ and $\boldsymbol{x}'$ and between all flavor pairs formed out of the $\mathscr{N}$ fermions. 
Strong correlations at a given site are described by \cite{Esterlis-2019,Inkof-2022,Hauck-2020,Wang2020b,Wang-2020a,Classen-2022,Choi-2022}
 \begin{eqnarray}
   \hat{H}_{\boldsymbol{x}}&=&-\sum_{i=1}^{\mathscr{N}}\sum_{\sigma=\pm}\mu \hat{c}_{i,\boldsymbol{x} ,\sigma}^{\dagger}\hat{c}_{i,\boldsymbol{x} ,\sigma}+\frac{1}{2}\sum_{k=1}^{\mathscr{M}}\left(\pi_{k\boldsymbol{x}}^{2}+\omega_{0}^{2}\phi_{k\boldsymbol{x}}^{2}\right) \nonumber \\
    &+&\frac{1}{\mathscr{N}}\sum_{\left \{i,j\right\}=1}^{\mathscr{N}}\sum_{\sigma=\pm}\sum_{k=1}^{\mathscr{M}}g_{ij,k}\hat{c}_{i,\boldsymbol{x},\sigma}^{\dagger}\hat{c}_{j,\boldsymbol{x},\sigma}\phi_{k\boldsymbol{x}},
\end{eqnarray}
In addition to fermions we have phonons, i.e.\@ scalar bosonic degrees of freedom $\phi_{k \boldsymbol{x}}$ with canonical momentum $\pi_{k\boldsymbol{x}}$.
$\left\{i,j\right\}=\left\{1\cdots \mathscr{N}\right\}$ refer to fermionic modes and $k=\left\{1\cdots \mathscr{M}\right\}$ to the phonon field. We consider the limit of large $\mathscr{N}$ and $\mathscr{M}$ with variable ratio $\mathscr{M}/\mathscr{N}$. 
The properties of the single-site model, in the replica-diagonal large-$\mathscr{N}$ ansatz and at particle-hole symmetry, were discussed in detail in Ref.\@ \onlinecite{Esterlis-2019}. The key finding was a self-tuned critical normal state at lowest temperatures, and a pairing state with a transition temperature $T_{\rm c}\propto g^2$, where $g$ is the typical value of the Gaussian-distributed coupling constants $g_{ij,k}$. Based on the analysis of similar purely fermionic models, one expects at finite $t_0$ a crossover to FL behavior below some energy scale $k_B T^*$; see below. The analysis of this model in the large-$\mathscr{N}$ limit is rather standard \cite{long-paper} and leads to the following saddle-point equations:
\begin{subequations}\label{eq:Eliashberg_eph:hop}
\begin{align}
\Sigma(i \omega_n)&=\bar{g}^2 k_B T \sum_{m=-\infty}^{+\infty} D(i\Omega_m) G(i \omega_n-i\Omega_m) \notag \\  &\textcolor{black}{+\frac{ t_0^2}{2} G(i \omega_n)}
\label{eq:Sigma}
\\
\Phi(i \omega_n)&=-\bar{g}^2 k_B T \sum_{m=-\infty}^{+\infty} F(i \omega_m) D(i \omega_n-i \omega_m) \notag \\  &\textcolor{black}{-\frac{t_0^2}{2} F(i \omega_n)}
 \label{eq:Phi}
\\
\Pi(i \Omega_n)&=-2 \bar{g}^2 k_B T \sum_{m=-\infty}^{+\infty}\left[ G(i \omega_m) G(i \omega_m+i \Omega_n)\right. \notag \\ & \left. -F(i \omega_m) F(i \omega_m+i \Omega_n)\right]
\label{eq:Polar}
\end{align}
\label{saddle_point_equations}
\end{subequations}
\begin{figure*}[t]
\includegraphics[width=0.9\textwidth]{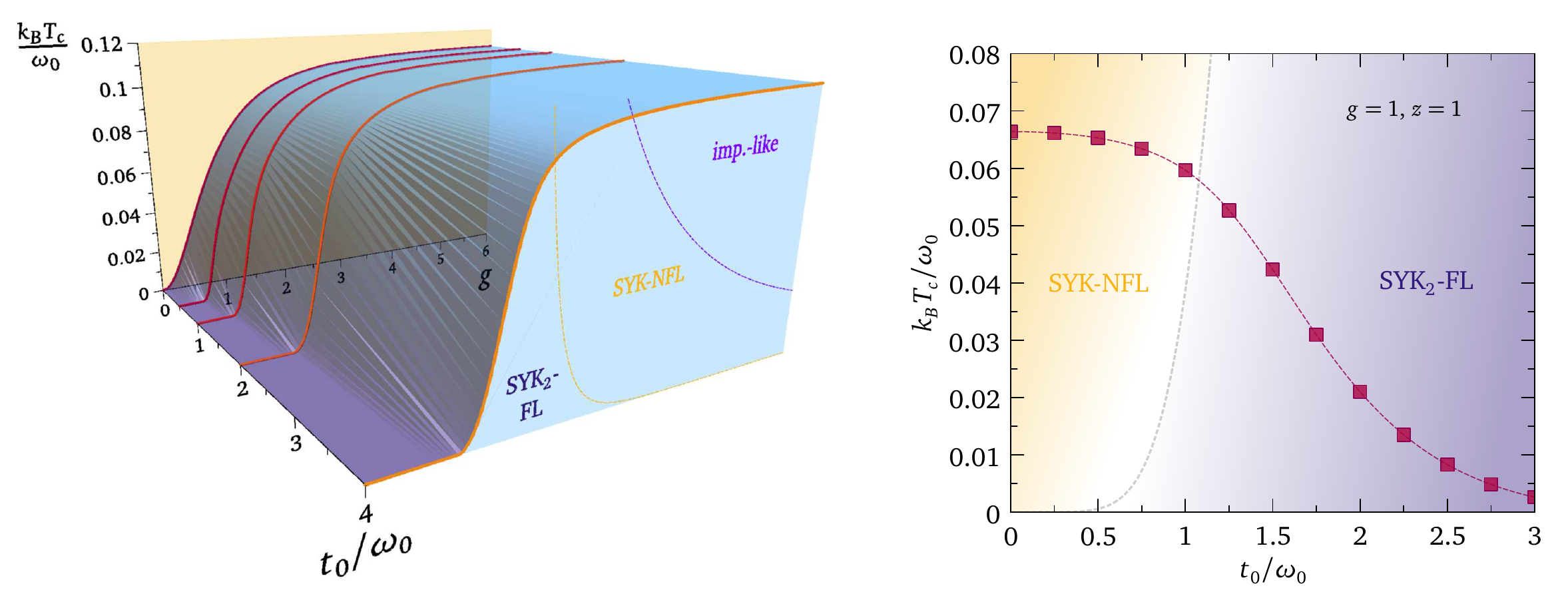}
\caption{\label{fig:Tc_3D} Superconducting transition temperature in the lattice model. Left panel: surface for $k_B T_c/\omega_0$ as a function of coupling $g$ and hopping $t_0$, showing the weak-coupling FL, NFL, and impurity-like regimes. Right panel: $k_B T_c/\omega_0$ at fixed coupling $g=1$, as a function of $t_0$, showing that $T_c$ is highest in the single-dot NFL regime.}
\end{figure*} 
Here, $\Sigma$ and $\Phi$ are the normal and anomalous fermionic self-energies of the Eliashberg formalism, while $G$ and $F$ are the corresponding propagators. $\Pi$ and $D$ are the bosonic self-energy and propagator, respectively. Eqs.\@ (\ref{eq:Eliashberg_eph:hop}) are supplemented by the Dyson equations
\begin{subequations}\label{eq:props}
\begin{align}
&
G(i \omega_n)=\frac{i \omega_n-\mu+\Sigma^{*}(i \omega_n)}{\mathscr{D}(\mu,i \omega_n)},
\\
& F(i \omega_n)=\frac{\Phi(i \omega_n)}{\mathscr{D}(\mu,i \omega_n)},
\\
&
D(i\Omega_n)=\frac{1}{\Omega_n^2+\omega_0^2-\Pi(i \Omega_n)},
\end{align}
\end{subequations}
where $\mathscr{D}(\mu,i \omega_n)=\left[i \omega_n+\mu-\Sigma(i\omega_n)\right]\left[i \omega_n-\mu+\Sigma^{*}(i\omega_n)\right]$ $-\left|\Phi(i \omega_n)\right|^2$ \cite{long-paper,Marsiglio-2008,Marsiglio-2020,Berthod-2018}.
In what follows we discuss the solution of the set (\ref{eq:Eliashberg_eph:hop}), (\ref{eq:props}), which determines quantities like the optical conductivity, the condensation energy, and the superfluid stiffness \footnote{In the following, we assume coordination $z=1$, $\mathscr{N}=\mathscr{M}$, and particle-hole symmetry ($\mu=0$) corresponding to a filling of 1 per site and flavor. Such filling is representative of a wide range of carrier concentrations except close to the band edge \cite{Wang-2020b}.}.

\begin{figure*}[t]
\includegraphics[width=0.99\textwidth]{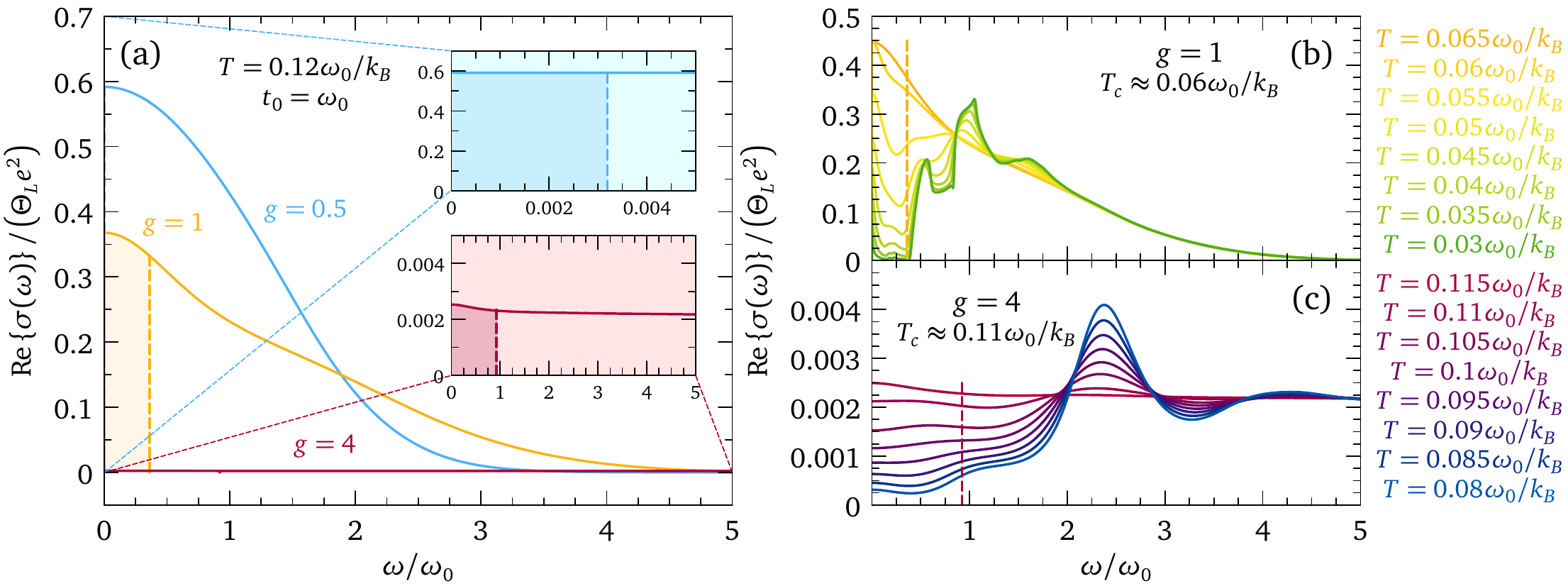}
\caption{\label{fig:conduct_SYK_T0} Real part of the optical conductivity $\sigma(\omega)$ in the SYK$_2$-FL, SYK-NFL, and impurity-like regimes, from the numerical solution of Eqs.\@ (\ref{eq:Eliashberg_eph:hop}) analytically continued to the real axis $\omega+i 0^+$, for $t_0=\omega_0$. Dashed vertical lines mark $\omega=2 \Delta_0$, with $\Delta_0$ gap at $T=0$ and $\omega=0$ estimated from Eqs.\@ (\ref{eq:Eliashberg_eph:hop}). (a) Normal-state conductivity $\sigma(\omega)/(\Theta_L e^2)$ at $k_B T=0.12 \omega_0$. $g=\left\{0.5,1,4\right\}$ give the light-blue, gold, and red curves, respectively. The shaded area highlights the spectral weight at $\omega\leq 2 \Delta_0$. The insets zoom on the SYK$_2$-FL and impurity-like regimes. (b) Temperature evolution of the SYK-NFL conductivity for $g=1$. (c) Temperature evolution of the impurity-like conductivity for $g=4$. }
\end{figure*}

{\em Normal state:} The behavior in the normal state is displayed in Figs.\@ \ref{fig:crossovers} and \ref{fig:conduct_SYK_T0}(a).  Fig.\@ \ref{fig:crossovers}(a) shows the phase diagram for uncoupled Yukawa-SYK dots. This is the behavior already discussed in Ref.\@ \onlinecite{Esterlis-2019}. Depending on temperature and the dimensionless coupling constant $g=\bar{g}/\omega_0^{3/2}$ there are three distinct normal-state regimes. At small $g$ and high temperature $k_B T\gg  g^2 \omega_0$ fermions form a classical gas. Interaction effects become important below the temperature $g^2 \omega_0$. At lowest $k_B T<\omega_0/g^2$, fermions and bosons form a strongly-coupled critical fluid, governed by a single universal exponent $G(i\omega)\propto |\omega|^{2\Delta-1}$ and $D(i\omega)\propto |\omega|^{1-4\Delta}$, where $1/4<\Delta <1/2$ depends on the ratio $N/M$. This SYK-NFL regime is similar to what is found for purely fermionic SYK models, where $G(i\omega)\propto |\omega|^{-1/2}$ for a four-fermion interaction. Once $g>1$, an intermediate impurity-like regime emerges, where bosons are extremely soft but sharp excitations, while fermions behave almost like in a disordered system with self-energy $\Sigma(i\omega)\sim -i {\rm sign}(\omega) g^2 \omega_0$, i.e.\@, with large scattering rate. 
In Fig.~\ref{fig:crossovers}(b,c) we show how the phase diagram changes as one allows for coherent single-particle hopping between SYK dots. The effect of hopping is straightforward for the classical gas. Once $k_B T<t_0$ a degenerate disordered Fermi gas forms. The behavior is identical to the so-called SYK$_{q=2}$  state, where $q$ is the number of fermion operators in the Hamiltonian \cite{Maldacena-2016a,Song-2017,Chowdhury-2022}. In the SYK-NFL state, the crossover scale is significantly reduced. Balancing the self energy of the SYK-NFL state and of the SYK$_{q=2}$ Fermi gas/Fermi liquid, one finds $k_B T^*\sim \left(t_0 g^{-4 \Delta}\right)^{1/(1-2\Delta)}$ for the crossover scale \cite{Esterlis-2019}, as confirmed numerically \cite{long-paper}. Further increasing hopping also pushes the crossover between the FL and impurity-like regimes to higher $g$, while it constrains the critical SYK state to an increasingly smaller window of parameters. In Fig.\@ \ref{fig:conduct_SYK_T0}(a) we show the normal-state optical conductivity for the SYK$_{2}$-FL, the SYK-NFL, and the impurity-like regimes, calculated from the electromagnetic kernel in linear response \footnote{For details on the derivation of the optical conductivity, please consult Appendix G of the companion paper \cite{long-paper}}. In particular, we observe a change in the frequency dependence between the former two regimes, with Drude-like behavior for small $g$ and quantum critical decay for $g\sim1$. The conductivity in the impurity-like regime at $g=4$ drops dramatically, given the large fermionic scattering rates. To summarize, our rather simple model allows for a rich normal-state behavior with FL, quantum-critical, and strongly incoherent behavior. In the next step we analyze the corresponding behavior of the superconducting state.
\begin{figure*}[t]
\includegraphics[width=0.9\textwidth]{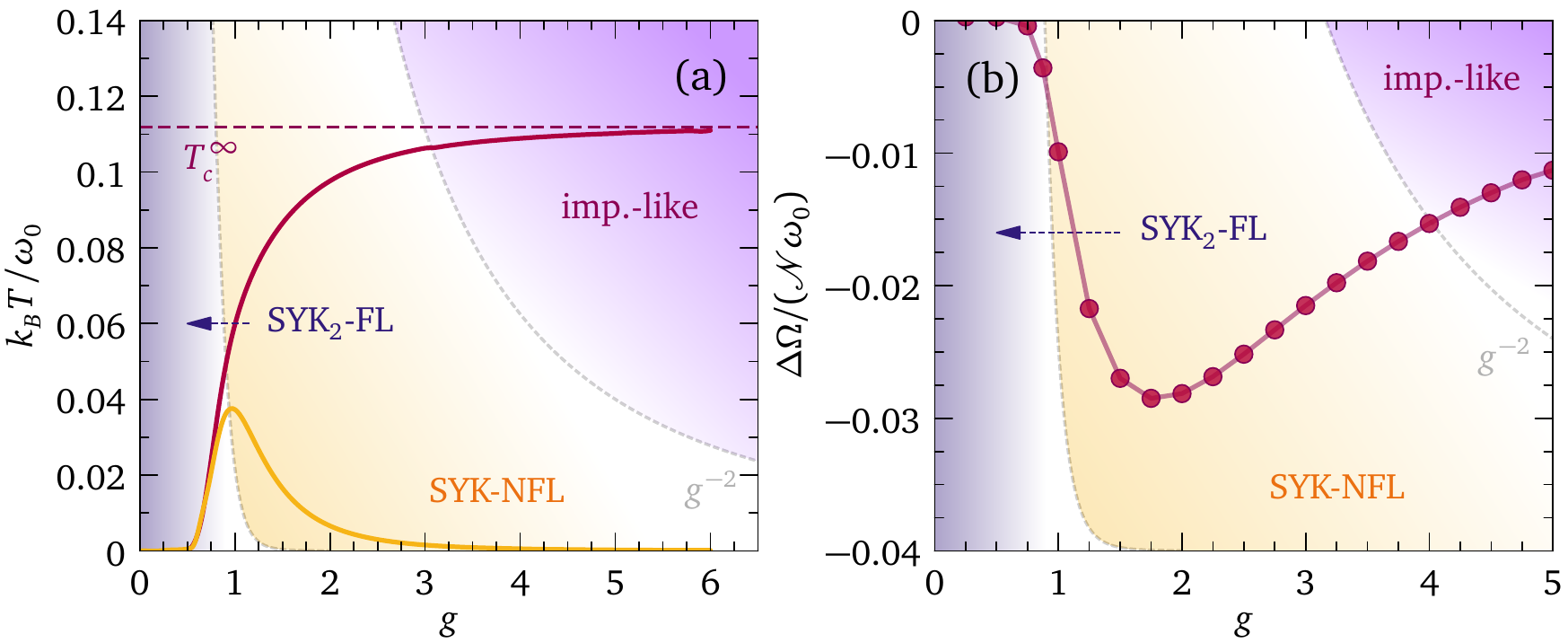}
\caption{\label{fig:pseudogap} Phase stiffness and condensation energy for $t_0=\omega_0$. (a) Normalized superfluid stiffness $\pi \rho_S/(\Theta_L e^\gamma \omega_0)$ at $k_B T=0.005 \omega_0$ (gold curve) and critical temperature $k_B T_{\rm c}/\omega_0$ (red curve) as a function of coupling $g$; $\Theta_L=\mathscr{N}/(a \hbar^2)$ where $a$ is the lattice constant \cite{long-paper}. (b) Condensation energy per fermion flavor $\Delta \Omega/(\mathscr{N} \omega_0)$ as a function of $g$, at $k_B T=0.01 \omega_0$.}
\end{figure*}

{\em Superconducting transition temperature:} In Fig.\@ \ref{fig:Tc_3D} we show the variation of the superconducting transition temperature $T_c$ with coupling constant $g$ and coherent hopping amplitude $t_0$. While the strong-coupling behavior is weakly affected by the single particle hopping, the weak- and intermediate-coupling behaviors change dramatically.
Pairing in the SYK-NFL regime was discussed in Ref.\@ \onlinecite{Esterlis-2019}: one finds that at small $g$ the transition temperature varies as $k_B T_{\rm c}\sim g^2 \omega_0$. This behavior is caused by the singular pairing interaction of NFL fermions, as mediated by the exchange of bosons $D(i\omega)\propto |\omega|^{1-4\Delta}$.  
The situation changes when a FL forms below $T^*$. We find that then Eqs.\@ (\ref{eq:Eliashberg_eph:hop}) lead to the established logarithmic Cooper instability of the SYK$_2$ regime, yet with a somewhat different origin than the standard BCS mechanism. The linearized version of the gap equation (\ref{eq:Phi})  can be written as 
\begin{equation}\label{eq:lin_gap_GG}
\tilde{\Phi}(i \omega_n)=\bar{g}^2 k_B T/\omega_0^2 \sum_{m}\frac{P(i \omega_m)}{1-z t_0^2 P(\omega_m)/2} \tilde{\Phi}(i \omega_m),
\end{equation}
where $P(i \omega_m)=G(i \omega_m)G(-i \omega_m)$ is the product of the propagators of the paired fermions. Using the analytical expression for $G(i \omega_m)$ in Fermi-liquid regime, and expanding the term under the sum for small $\omega_m$, one realizes that the leading-order term scales as $1/\omega_m$, which yields a logarithmic Cooper instability \cite{long-paper}. Thus, we find the asymptotically exact BCS weak-coupling expression $k_B T_{\rm c}=2 e^{\gamma}/\pi \omega_0 e^{-1/\bar{\lambda}}$, with $\gamma$ Euler-Mascheroni constant and with and effective coupling constant 
\begin{equation}
\bar{\lambda}=\frac{\sqrt{2}}{\pi} g^2 \omega_0/ t_0.
\end{equation}
Notice, the instability does not come from the term $P(i \omega_m)$ at the numerator of Eq.\@ (\ref{eq:lin_gap_GG}), as would be the case for textbook BCS pairing, but it is generated by the term $1- z t_0^2 P(i \omega_m)/2$ in the denominator \cite{long-paper,Li-2023}. 
 
Comparing the the expressions for the transition temperatures at small $g$, $T_c$ of a critical, NFL state is large compared to that what follows from the usual Cooper instability. The lack of coherency of the NFL state, which at first glance weakens the instability, is more than compensated by the singular pairing strength due to the exchange of quantum critical bosons. This behavior is illustrated in the right panel of Fig.~\ref{fig:Tc_3D} where we show $T_c$ as function of $t_0$ at fixed $g$. This is reminiscent of the emergence of a superconducting dome that forms around a quantum critical point. The rapid drop in $T_c$ coincides with the crossover from the quantum critical to the Fermi-liquid regime.

{\em Superfluid stiffness:} The robustness of a $d$-dimensional superconducting state against fluctuations of the phase of the order parameter is determined by the superfluid stiffness 
\begin{equation}
\rho_S=L^{2-d}\left. \frac{\partial^2 F(\theta)}{\partial \theta^2} \right|_{\theta \rightarrow 0}
\end{equation} 
with free energy as function of an applied phase twist $F(\theta)$ \cite{Fisher-1973,Taylor-2006}, and $L$ linear sample dimension. For Galilean-invariant systems it holds at $T=0$ that $\rho_S=n/m$, with particle number $n$ and mass $m$, respectively \cite{Leggett-1998,Leggett-2006quantum}. This implies negligible phase fluctuations as long as $k_B T_{\rm c}\ll E_{\rm F}$ with Fermi energy $E_{\rm F}$. For disordered superconductors with impurity scattering rate $\tau^{-1}$ it holds in the limit $\Delta \ll \tau^{-1} \ll E_{\rm F}$ that the ground-state stiffness is reduced to $\rho_S/(n/m)=\pi \tau \Delta$\cite{Abrikosov-2012meth}, i.e. it scales with the superconducting gap $\Delta$, or equivalently, with the transition temperature $T_c$.

In Fig.\@ \ref{fig:pseudogap}(a) we show the $g$ dependence of the stiffness at low temperatures for finite $t_0=\omega_0$. $\rho_S$ is displayed as an energy scale in units of $\Theta_L$ \cite{long-paper}. In the weak-coupling regime $\pi \rho_S/(\Theta_L e^\gamma)$ (gold curve) tracks perfectly $k_B T_{\rm c}$ (red curve), precisely as one expects for a disordered BCS superconductor. On the other hand, at large $g$, the stiffness decreases with increasing $g$ like $\rho_S \sim g^{-4}$, as the superconducting state that forms from an incoherent normal state becomes increasingly more fragile \cite{long-paper}. Such fragility could be reflected by an analysis of phase fluctuations as $1/\mathscr{N}$ corrections with respect to our saddle-point theory. $\rho_S$ is largest right at the crossover between the FL and NFL regimes. These trends can be understood from the coupling constant dependence of the optical conductivity shown in Fig.\@ \ref{fig:conduct_SYK_T0}. The stiffness can be estimated using the Ferrell-Glover-Tinkham (FGL) sum rule \cite{Ferrell-1958,Tinkham-1959}
\begin{equation}\label{eq:FGL_sum_rule}
\rho_S=\frac{2}{\pi e^2} \int_{0^+}^\infty d\omega \left( \mathrm{Re}\left\{\sigma_{\rm ns}(\omega)\right\}-\mathrm{Re}\left\{\sigma_{\rm sc}(\omega)\right\}\right),
\end{equation}
where $\mathrm{Re}\left\{\sigma_{\rm ns}(\omega)\right\}$ and $\mathrm{Re}\left\{\sigma_{\rm sc}(\omega)\right\}$ are the real parts of the optical conductivities of the normal and superconducting states, respectively. While the FGL sum rule assumes that the total optical weight is unchanged, it allows for a qualitative understanding of the trends that we see in the $\rho_S$ shown in Fig.\@ \ref{fig:pseudogap}(a). The stiffness of the weak-coupling disordered FL is small because of the small superconducting gap, compared to the scattering rate. Hence only a small portion of the total spectral weight is being transformed into the $\omega=0$ $\delta$-function $\mathrm{Re}\left\{\sigma_{\rm sc}(\omega)\right\}=\pi \rho_S \delta(\omega)$.
This portion is qualitatively estimated by the light-blue shaded area in the inset of Fig.\@ \ref{fig:conduct_SYK_T0}(a), where the dashed line corresponds to $2 \Delta_0$ for $g=0.5$, $\Delta_0$ being the $T=0$ static superconducting gap. In the DC limit and at $T=0$, we obtain $\mathrm{Re}\left\{\sigma_{\rm ns}(0)\right\} /\Theta_L=2/\pi$, so that from Eq.\@ (\ref{eq:FGL_sum_rule}), $\rho_S\approx \frac{2}{\pi} \mathrm{Re}\left\{\sigma_{\rm ns}(\omega)\right\} 2 \Delta_0 \approx 0.81 \Delta_0$. This estimation is fully consistent with a direct zero-temperature calculation in FL regime yielding $\rho_S/\Theta_L= \Delta_0$ \cite{long-paper}.
On the other hand, in the incoherent regime at large $g$ the conductivity is small: precisely, in the DC limit we have $\mathrm{Re}\left\{\sigma_{\rm ns}(0)\right\} /\Theta_L=4/(\pi \Omega_0^2)$, where $\Omega_0=16 g^2 \omega_0/(3\pi)$ \cite{Esterlis-2019}. Hence, a small amount of weight is transferred despite a large pairing gap, and $\rho_S \sim g^{-4} \Delta_0$, as confirmed by the exact $T=0$ result $\rho_S/\Theta_L=2 z t_0^2/\Omega_0^2 \Delta_0$ \cite{long-paper}. This effect is highlighted by the red shaded area in the inset of Fig.\@ \ref{fig:conduct_SYK_T0}(a), delimited by $2 \Delta_0$ for $g=4$ (dashed line). The sweet spot occurs at the crossover, where the conductivity is still sizable and the pairing gap already large -- see gold-shaded area and dashed line for $g=1$ in Fig.\@ \ref{fig:conduct_SYK_T0}(a) -- explaining the maximum in the stiffness. 

{\em Superconducting optical conductivity:} Our arguments on the low-frequency shift of the spectral weight, according to Eq.\@ (\ref{eq:FGL_sum_rule}), are substantiated by the explicit calculation of the conductivity in the superconducting state. Figs.\@ \ref{fig:conduct_SYK_T0}(b,c) show the frequency dependence of $\mathrm{Re}\left\{\sigma(\omega)\right\}$ for $t_0=\omega_0$, in the SYK-NFL and impurity-like regimes respectively, at temperatures $T<T_c$. We observe the progressive depletion of spectral weight below $2 \Delta_0$ (dashed vertical lines), accompanied by shakeoff peaks at $\omega>2 \Delta_0$, due to the strongly interacting nature of the Cooper-pair fluid, and caused by quasiparticle self-trapping into the fluctuating pairing field. Such structures reflect analogous peaks in the fermionic and bosonic spectral functions \cite{Esterlis-2019,Li-2023,long-paper}. Interestingly, the highest-amplitude coherence peaks occur at $\omega>4 \Delta_0$ in the SYK-NFL and impurity-like phases. 

{\em Condensation energy:} The robust pairing state at the FL to NFL crossover can also be deduced from the condensation energy $\Delta \Omega=\Omega^{\rm sc}-\Omega^{\rm ns}$, where $\Omega^{\rm sc,ns}$ stand for the grand potential of the normal and superconducting state, respectively. Fig.\@ \ref{fig:pseudogap}(b) shows the coupling evolution of $\Delta \Omega$ from the exact numerical solution of Eqs.\@ (\ref{eq:Eliashberg_eph:hop}), for $t_0=\omega_0$.
$\Delta \Omega$ is largest very near where the stiffness is largest. It decreases at small and large coupling, because of the small $T_c$ in the former case and because of the small weight of the Bogoliubov quasiparticles in the latter.

{\em Summary:} We analyzed a model of strongly-interacting, critical Yukawa SYK sites, coupled via a random coherent single particle hopping. Hopping gives rise to a crossover temperature $T^*$ that separates quantum critical, NFL behavior at intermediate temperatures and FL behavior at lowest temperatures. Pairing of the NFL state is significantly more efficient than the one due to the conventional Cooper instability, accounting for the frequently observed maximum in the transition temperature at quantum critical points. The superfluid stiffness and the condensation energy are largest right at the crossover between FL and NFL behavior. At large coupling constant, the weight of the Bogoliubov quasiparticles, the stiffness and the condensation energy are all correlated, similar to what has been reported for the behavior in underdoped cuprates \cite{long-paper}. These trends are also descerned in the optical response, and are believed to be governed by generic principles that characterize pairing of systems without quasiparticles.

D.\@ V.\@ acknowledges partial support by the Swiss National Science Foundation (SNSF) through the SNSF Early Postdoc.Mobility Grant P2GEP2\_181450, and by the Mafalda cluster of the University of Geneva, on which some of the calculations were performed. G.\@ A.\@ I.\@ and J.\@ S.\@ were supported by the Deutsche Forschungsgemeinschaft (DFG, German Research Foundation) - TRR 288 - 422213477 Elasto-Q-Mat (project A07).   

%

\end{document}